\documentclass[prb,a4paper,twocolumn,floatfix,showpacs,showkeys,amsmath,amssymb,nobibnotes,altaffilletter]{revtex4}

\usepackage{graphicx}% Include figure files
\usepackage{pslatex}
\usepackage{xspace}
\usepackage{subfigure}

\newcommand{\pht}{poly(3-hexyl thiophene)\xspace}
\newcommand{\pcbm}{[6,6]-phenyl-C$_{61}$ butyric acid methyl ester\xspace}
\newcommand{\degree}{$^{\circ}$}

\newcommand{\beq}{\begin{equation}}
\newcommand{\eeq}{\end{equation}}

\begin{document}
\preprint{}

%\title{Delayed Bimolecular Polaron Recombination in Polymer--Fullerene Solar Cells}
%\title{High Order Charge Carrier Decay in Polymer--Fullerene Solar Cells}
\title{Charge Carrier Concentration and Temperature Dependent Recombination in Polymer--Fullerene Solar Cells}

\author{A.~Foertig$^{1,2}$}
\author{A.~Baumann$^1$}
\author{D.~Rauh$^{1,2}$}
\author{V.~Dyakonov$^{1,2}$}\email{dyakonov@physik.uni-wuerzburg.de}
\author{C.~Deibel$^1$}\email{deibel@physik.uni-wuerzburg.de}

\affiliation{$^1$Experimental Physics VI, Julius-Maximilians-University of W{\"u}rzburg, 97074 W{\"u}rzburg, Germany}
\affiliation{$^2$Bavarian Centre for Applied Energy Research (ZAE Bayern), 97074 W{\"u}rzburg, Germany}

\date{\today}

\begin{abstract}

We performed temperature dependent transient photovoltage and photocurrent measurements on \pht :\pcbm bulk heterojuction solar cells. We found a strongly charge carrier concentration and temperature dependent Langevin recombination prefactor. The observed recombination mechanism is discussed in terms of bimolecular recombination. The experimental results were compared with charge carrier extraction by linearly increasing voltage (photo-CELIV) measurements done on the same blend system. We explain the charge carrier dynamics, following an apparent order larger than two, by dynamic trapping of charges in the tail states of the gaussian density of states.

\end{abstract}



\maketitle
As power conversion efficiencies of solution processed organic solar cells have reached $6\%$, these low cost photovoltaic devices attract more and more interest.\cite{green2008,park2009} To achieve further improvements a better understanding of the fundamental loss processes, such as bimolecular charge carrier recombination, is needed. A suitable technique to investigate organic solar cells under operating conditions in terms of charge carrier lifetimes and densities are transient photovoltage (TPV) and transient photocurrent (TPC) which were recently applied to determine charge carrier decay in polythiophene:fullerene solar cells.\cite{shuttle2008} An experimentally observed third order dependence on charge carrier density was interpreted in previous reports with a bimolecular recombination in combination with a carrier concentration dependent prefactor.~\cite{shuttle2008,deibel2008b}  Therefore, we investigated charge carrier loss processes in \pht (P3HT):\pcbm (PCBM) bulk heterojunction solar cells using the complementary experimental methods TPV, TPC and photo-CELIV. We found a strongly temperature and carrier concentration dependent polaron recombination prefactor.\\

The investigations were performed on a bulk heterojunction solar cell of a 1:1 blend of a 90 nm thick film of regioregular P3HT (Rieke Metals) and PCBM (Solenne b.v.). The blend was spin coated from a solution of 30 mg/ml chlorobenzene on a poly(3,4-ethylendioxythiophene):polystyrolsulfonate coated indium tin oxide/glass substrate. After an annealing step of 10 min at 130\degree{}C, Ca/Al contacts were evaporated thermally. The samples were processed in a nitrogen glovebox and an attached thermal vacuum evaporation chamber, also described eleswhere.\cite{deibel2008b} By using this recipe we obtained power conversion efficiencies (uncertified) in the range between $3-4\%$, after a correction related to the spectral characteristics of the lamp consistent with external quantum efficiency measurements.\\

TPV and TPC were applied using a 10W high power white light LED (Cree), including focussing optics as bias light source. A nitrogen laser pulse (Spectra Physics VSL-337 ND-S, $\lambda$=500~nm by dye unit, pulse duration 5~ns, energy density 500~$\mu$J/cm$^2$) was applied to change either photovoltage  or photocurrent of the device. Its illumination intensity was attenuated by neutral density filters in order to ensure a small perturbation. Voltage transients after laser excitation were aquired by a digital storage oscilloscope (Tektronix, 1~GHz band width, sample rate 4~GS/s) in order to determine the small pertubation carrier lifetime $\tau_{\Delta n}$ under open circuit conditions. In addition, from photocurrent transients at short circuit conditions we were able to determine the charge carrier concentration $n$ in analogy to Ref.~3. Photo-CELIV measurements were done in the same setup as described previously.\cite{deibel2008b} By varying the delay time between the exciting laser pulse and the triangular voltage extraction pulse, the time dependent carrier concentration and mobility can be determined simultaneously.\cite{juska2000} A Helium closed-cycle cryostate with contact gas was used to vary the temperature. \\

The continuity equation
\beq
	\frac{dn}{dt}=G-R_{mr}-R_{br} \quad{,}
	\label{eq:Recrate}
\eeq
with the polaron generation rate $G$ (including the efficient exciton dissociation as well as geminate losses) respectively describes the charge carrier dynamics within an organic solar cell. $R_{mr}$ and $R_{br}$ are the non-geminate monomolecular and bimolecular recombination rates. Monomolecular recombination is defined as one mobile charge carrier recombining with an immobile trapped charge.
For disordered materials with low carrier mobilities in the range of $10^{-8}$~m$^2$/Vs, Langevin theory is often considered to describe polaron losses most suitably, where bimolecular recombination is considered as a process of two charges finding each other by diffusion.\cite{langevin1903,pope1999} Therefore, the Langevin recombination rate
\beq
	R_{br}=k_{br}np
\label{eq:BR}
\eeq
is related to the sum of electron and hole mobility $\mu$ by the Langevin prefactor
\beq
	k_{br}=\frac{q}{\epsilon}\mu \quad{.}
\label{eq:prefactor}
\eeq
$n$ and $p$ are the negative and positive polaron densities, $q$ is the elementary charge and $\epsilon$ the material dielectric constant. Previous studies have shown that an additional factor between $10^{-1}$ and $10^{-4}$ compared to Langevin's original theory is needed to describe the charge carrier dynamics appropriately.\cite{juska2006,deibel2008b,deibel2009}\\

The small pertubation carrier lifetime $\tau_{\Delta n}$ extracted from TPV can be expressed as
\beq
	\tau_{\Delta n}=\tau_{\Delta n_0}\left(\frac{n_0}{n}\right)^{\lambda}\quad{.}
	\label{tau}
\eeq
$\tau_{\Delta n_0}$, $n_0$ and $\lambda=\beta/\gamma$ are temperature dependent parameters which we determined experimentally. The details of the methods can be found in Shuttle et al.~\cite{shuttle2008} By using Eq.~(5) from Ref.~3 and assuming bimolecular recombination and $n=p$, the total carrier dynamics can be described as
\beq
	\frac{dn}{dt} \approx - k_\lambda n^{\lambda+1} \quad{,}
	\label{kbr}
\eeq
with $k_\lambda = 1/((1+\lambda)\tau_{\Delta n_0} n_0^{\lambda})$~. Defining
\beq
	\frac{dn}{dt} = -k_{br} n^2 \quad{,}
	\label{kbr_def}
\eeq
and using $k_{br_0}(T)=1/(\tau_{\Delta n_0}n_0)$, the bimolecular Langevin recombination prefactor becomes
\beq
	k_{br}(T,P_L)=\frac{k_{br_0}(T)}{1+\lambda(T)}\left(\frac{n(T,P_L)}{n_0(T)}\right)^{\lambda(T)-1}\quad{,}
	\label{kbrcalc}
\eeq
where $P_L$ denotes the light intensity.\\

\begin{figure}[tb]
	\includegraphics[width=7cm]{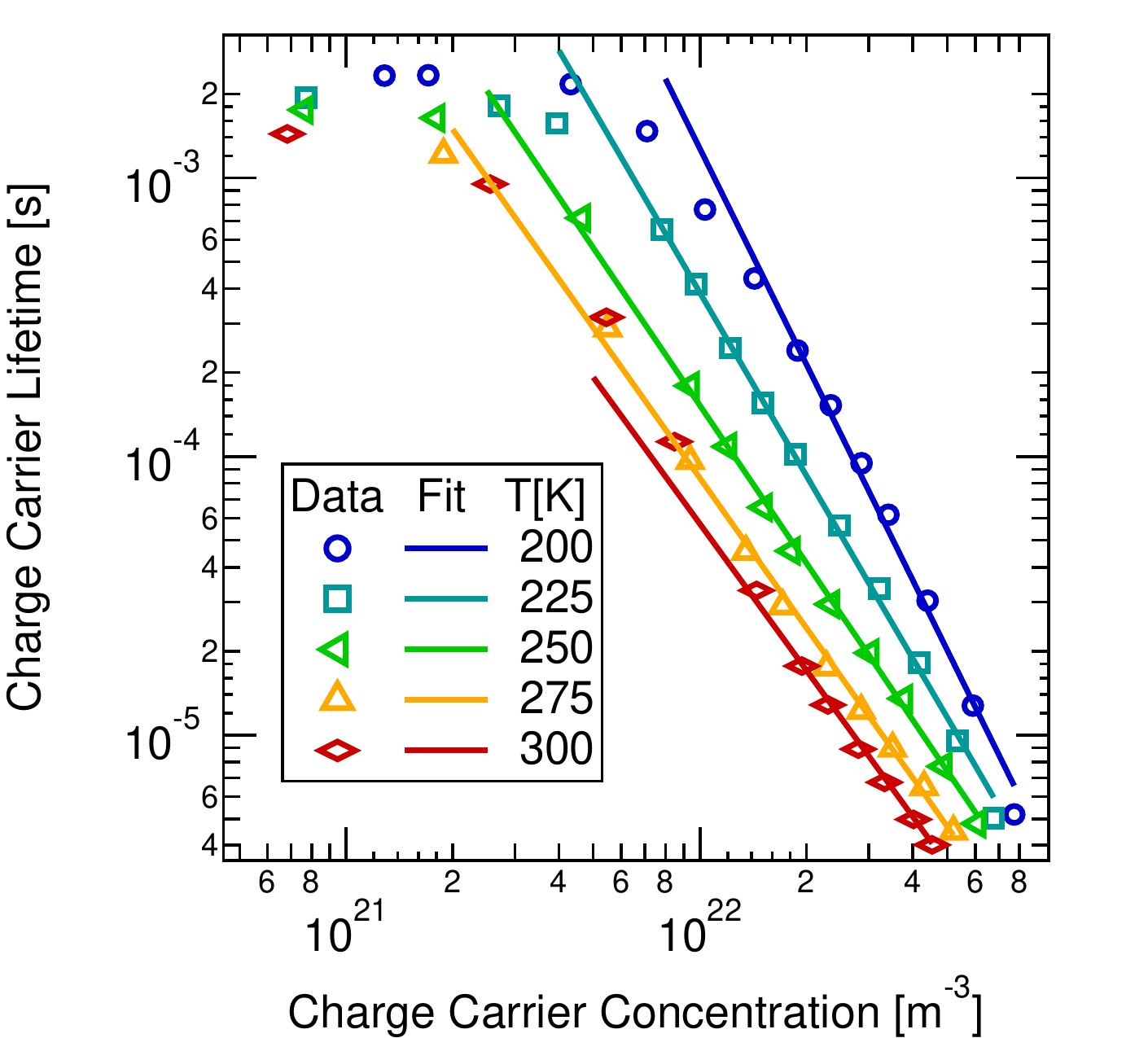}
	\caption{Small perturbation charge carrier lifetime as function of charge carrier concentration at different temperatures.}
	 \label{fig:fig1}
\end{figure}

In Fig.~\ref{fig:fig1}, the experimentally determined small pertubation charge carrier lifetime $\tau_{\Delta n}$ in dependence on charge carrier density is shown for various temperatures. Notable is an almost temperature independent value of $\tau_{\Delta n}$ for charge carrier densities lower than 3$\cdot$10$^{21}$~m$^{-3}$. A limitation by the dielectric relaxation time\cite{juska2000} can be excluded, as it is proportional to $(n\mu)^{-1}$, and thus expected to be strongly temperature dependent. Above $n=3\cdot$10$^{21}$~m$^{-3}$, the lifetime $\tau_{\Delta n}$ exhibits a nonlinear decrease with $n$. In accordance with Eq.~(\ref{eq:Recrate}) this indicates two distinct mechanisms taking place, a monomolecular process with almost temperature independent lifetime $\tau_{\Delta n_\text{mr}}$ for low carrier densities and a bimolecular process with $\tau_{\Delta n_\text{br}}$(n) for higher densities. 
Thus, the effective small perturbation carrier lifetime is given as
\beq
\tau_{\Delta n}(n)=\left(\frac{1}{\tau_{\Delta n_\text{mr}}}+\frac{1}{\tau_{\Delta n_\text{br}}(n)}\right)^{-1}\quad{.}
\eeq
In the following we focus on the bimolecular process to determine the recombination prefactor and its charge carrier concentration and temperature dependence.

\begin{table}[tb]
\centering
\caption{Temperature dependence of the parameter $\lambda$.}
\begin{tabular}{c||c|c|c|c|c}
$T[K]$	 &300& 275& 250& 225& 200\\
\hline
$\lambda$&1.75& 1.79& 1.88& 2.18& 2.58\\
\end{tabular}
\label{tab:lambda}
\end{table}

The effective charge carrier decay order given in Eq.~(\ref{kbr}), corresponds to $\lambda+1$. The parameter $\lambda$ increases from 1.75 to 2.58 with decreasing temperature, as shown in Tab.~\ref{tab:lambda}. Although the apparent order of the recombination process is larger than two, we expect bimolecular decay for the reasons discussed below.
\begin{figure}
 	\includegraphics[width=7cm]{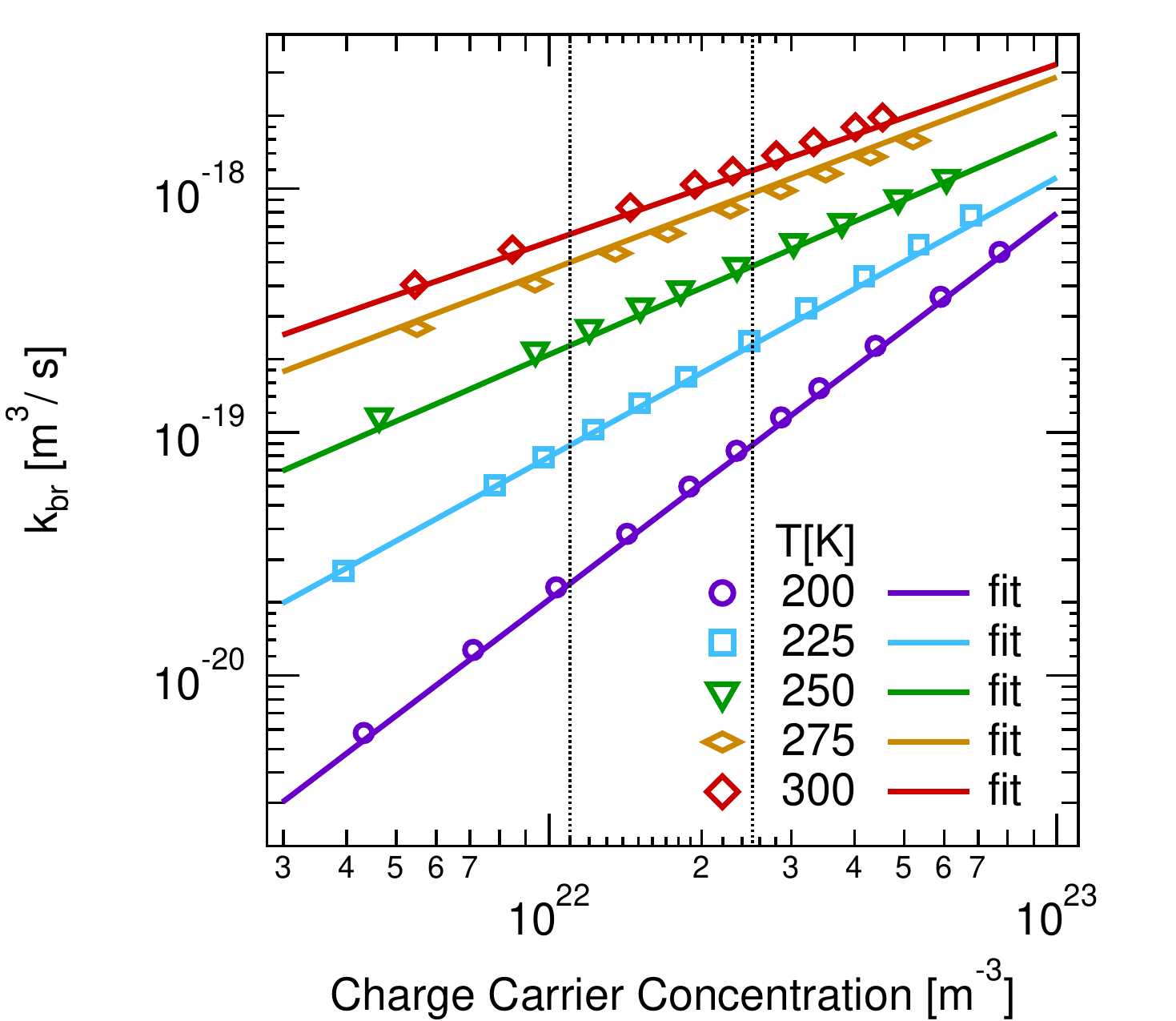}
 	\caption{Carrier concentration dependent bimolecular recombination prefactor $k_{br}(n)$ for several temperatures. Vertical dashed lines indicate charge carrier density values used to study the temperature dependence of $k_{br}$ in Fig.~\ref{fig:fig3}.}
 	 \label{fig:fig2}
\end{figure}

We apply Eq.~(\ref{kbrcalc}) to determine the recombination prefactor $k_{br}$. It rises steeply with increasing charge carrier concentration $n$, as shown in Fig.~\ref{fig:fig2}. 
\begin{figure}
	\includegraphics[width=7cm]{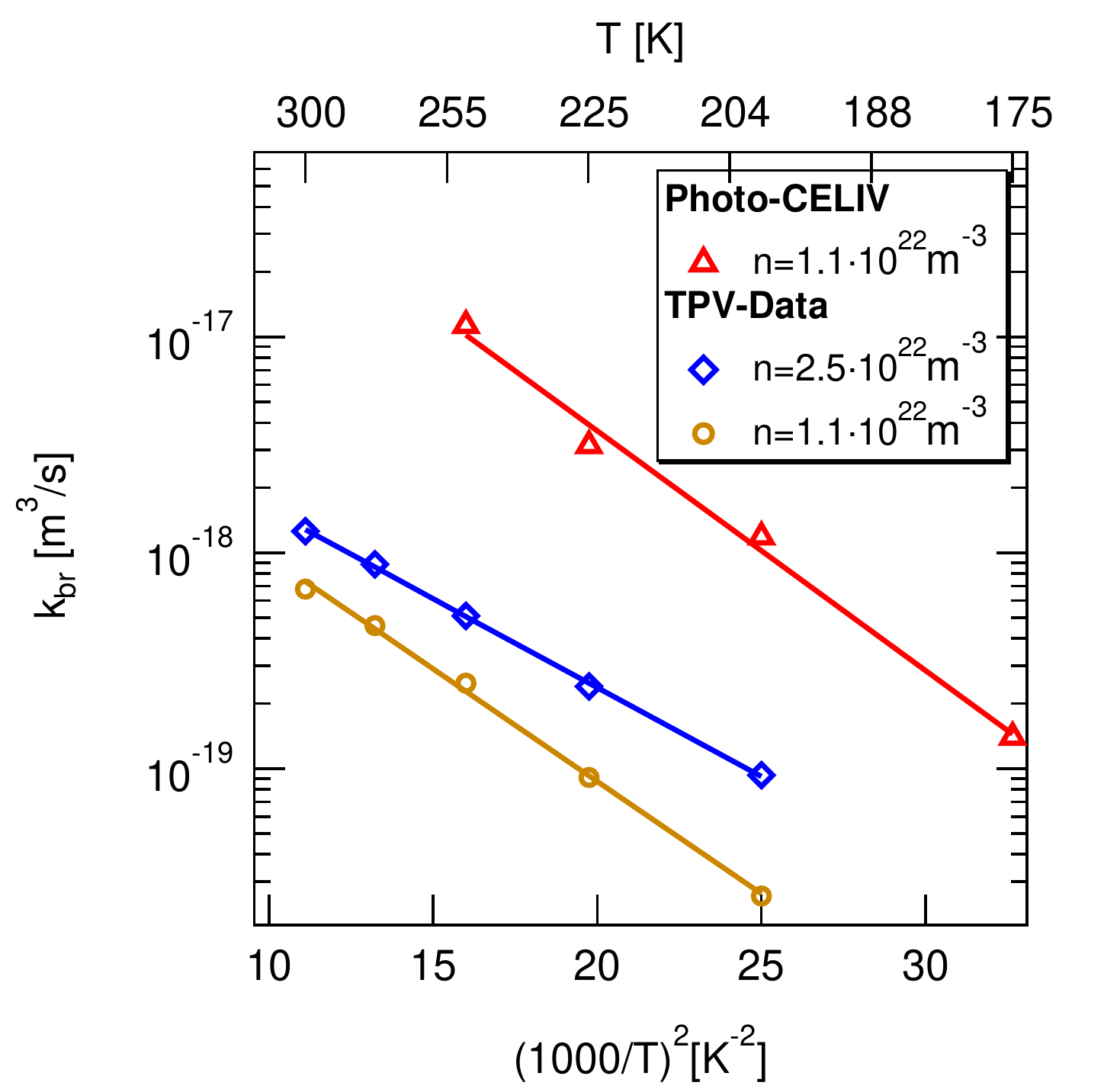}
	\caption{Temperature dependence of the bimolecular recombination prefactor for certain carrier concentrations from TPV and photo-CELIV measurements.}
	 \label{fig:fig3}
\end{figure}
In order to clarify the temperature dependence of $k_{br}$, we calculated its cross sections at distinct carrier concentrations, $n=1.1\cdot 10^{22}$~m$^{-3}$ and $2.5\cdot 10^{22}$~m$^{-3}$, for all temperatures available. The result is presented in Fig.~\ref{fig:fig3} and complemented by bimolecular recombination prefactors determined by photo-CELIV measurements on the same sample as this is a more established technique in the field of organic photovoltaic. We point out that the absolute values of $k_{br}$ from photo-CELIV measurements are one order of magnitude higher compared to the corresponding TPV measurement. We attribute this discrepancy to the different measurement conditions: TPV and TPC are measured under constant illumination with small pertubation in the quasi-equilibrium region, whereas photo-CELIV experiments were performed using a short intense laser pulse in the non-equilibrium without bias light. The fits of the data show clearly that the temperature dependence of the recombination factor can be described by $k_{br} \propto \exp (-(\sigma^*/kT)^2)$, with $\sigma^*=42$~meV for $n=1.1\cdot 10^{22}$~m$^{-3}$ in both experiments. The analog functional dependence to the Gaussian disorder model is striking.\cite{bassler1993} This indicates that a major part of the temperature dependence of $k_{br}$ stems from the charge carrier mobility, in line with Eq.~(\ref{eq:prefactor}).\\

In view of our experimental results, and in accordance with previous publications,\cite{shuttle2008,deibel2008b} we interpret the charge carrier recombination in the P3HT:PCBM blend as bimolecular losses according to Langevin's theory.\cite{pope1999} The experimentally observed decay dynamics of almost third order at room temperature, and even higher order at lower temperatures, can be understood by the following considerations related to the charge transport being due to hopping between the localized states of a gaussian density of states.
The total carrier concentration $n$ can be represented by the sum of carriers in extended states, $n_c$, responsible for the conductivity, and trapped carriers $n_t$ in the tails of the gaussian density of states.\cite{rudenko1982,arkhipov1982} The ratio $n_c$/$n_t$ being smaller than 1 at low bias light intensities might explain the dominating monomolecular recombination process at lower charge carrier densities. Zaban et al. derived the dependence of the experimental carrier lifetime 
\beq
	\tau_n = \left( 1 + \frac{dn_t}{dn_c} \right) \tau_{n0} 
	\label{eq:tau_zaban}
\eeq
under quasi-steady state conditions---as found in TPV/TPC measurements--- for a material system with an exponential density of states.\cite{zaban2003} As this derivation is valid for carrier concentration dependent lifetime, Eq.~(\ref{eq:tau_zaban}) is directly relevant to bimolecular recombination as well.
The lifetime $\tau_{n0}$ is independent of the carrier concentrations, describing the trap-free case---in our case the classical bimolecular recombination with constant prefactor. The authors assigned the derivative $dn_t/dn_c$ to a delay factor for recombination due to intermittent trapping of free carriers in the tails of the density of states. This occurs already under the quasi-steady state conditions under bias illumination with additional trapping during the excess charge generation and recombination in a transient measurement. Trapped charges act only as immobile recombination partners but can be released at a later time---during the transient measurement or afterwards--- to participate again actively in the recombination process. \\

The principles of trapping and release of charge carriers under quasi-steady state\cite{rudenko1982} and non-equilibrium\cite{arkhipov1982} have already been described decades ago. For the expected case of strong trapping, $dn_\text{t}/dn_\text{c}\gg 1$. Thus, $n_t$ is a function of the free carrier concentration $n_c$, which explains the apparent carrier concentration dependence of the recombination prefactor  qualitatively. Consequently, the magnitude of the decay order, $\lambda+1$, is expected to depend strongly on temperature, as the release from a trap is a thermally activated process. And indeed, this interpretation is consistent with the carrier concentration dependence of $k_{br}$ becoming much weaker at high temperatures (Fig.~\ref{fig:fig2}): there, the effective disorder $\sigma/kT$ becomes smaller\cite{bassler1993} and all charges have sufficient thermal energy to render trapping irrelevant.\\

To conclude, we performed temperature dependent TPV, TPC and photo-CELIV measurements on P3HT:PCBM bulk heterojunction solar cells. We found both, a monomolecular and a bimolecular recombination mechanism for photogenerated polarons. The latter becomes dominant at carrier concentrations above $3\cdot$10$^{21}$~m$^{-3}$ and shows an apparent decay of an order exceeding the expected bimolecular case. We explain the resulting carrier concentration dependence of the bimolecular recombination prefactor $k_{br}$ by the influence of trapping of charges in the tail states of the gaussian density of states. This interpretation is consistent with the weaker carrier concentration dependence of $k_{br}$ at higher temperatures, where the effect of trapping and release due to disorder becomes negligible. Our findings have important implications on the understanding and modeling of organic solar cells.

\begin{acknowledgments}

The current work is supported by the Bundesministerium f{\"u}r Bildung und Forschung in the framework of the GREKOS project (Contract No. 03SF0356B).
A.B. thanks the Deutsche Bundesstiftung Umwelt for funding. V.D.'s work at the ZAE Bayern is financed by the Bavarian Ministry of Economic Affairs, Infrastructure, Transport and Technology. 

\end{acknowledgments}

%\bibliography{foertig2009}

\begin{thebibliography}{13}
\expandafter\ifx\csname natexlab\endcsname\relax\def\natexlab#1{#1}\fi
\expandafter\ifx\csname bibnamefont\endcsname\relax
  \def\bibnamefont#1{#1}\fi
\expandafter\ifx\csname bibfnamefont\endcsname\relax
  \def\bibfnamefont#1{#1}\fi
\expandafter\ifx\csname citenamefont\endcsname\relax
  \def\citenamefont#1{#1}\fi
\expandafter\ifx\csname url\endcsname\relax
  \def\url#1{\texttt{#1}}\fi
\expandafter\ifx\csname urlprefix\endcsname\relax\def\urlprefix{URL }\fi
\providecommand{\bibinfo}[2]{#2}
\providecommand{\eprint}[2][]{\url{#2}}

\bibitem[{\citenamefont{Green et~al.}(2008)\citenamefont{Green, Emery,
  Hishikawa, and Warta}}]{green2008}
\bibinfo{author}{\bibfnamefont{M.~A.} \bibnamefont{Green}},
  \bibinfo{author}{\bibfnamefont{K.}~\bibnamefont{Emery}},
  \bibinfo{author}{\bibfnamefont{Y.}~\bibnamefont{Hishikawa}},
  \bibnamefont{and} \bibinfo{author}{\bibfnamefont{W.}~\bibnamefont{Warta}},
  \bibinfo{journal}{Phys. Plasmas} \textbf{\bibinfo{volume}{17}}
  (\bibinfo{year}{(2008)}).

\bibitem[{\citenamefont{Park et~al.}(2009)\citenamefont{Park, Roy, Beaupre,
  Cho, Coates, Moon, Moses, Leclerc, Lee, and Heeger}}]{park2009}
\bibinfo{author}{\bibfnamefont{S.~H.} \bibnamefont{Park}},
  \bibinfo{author}{\bibfnamefont{A.}~\bibnamefont{Roy}},
  \bibinfo{author}{\bibfnamefont{S.}~\bibnamefont{Beaupre}},
  \bibinfo{author}{\bibfnamefont{S.}~\bibnamefont{Cho}},
  \bibinfo{author}{\bibfnamefont{N.}~\bibnamefont{Coates}},
  \bibinfo{author}{\bibfnamefont{J.~S.} \bibnamefont{Moon}},
  \bibinfo{author}{\bibfnamefont{D.}~\bibnamefont{Moses}},
  \bibinfo{author}{\bibfnamefont{M.}~\bibnamefont{Leclerc}},
  \bibinfo{author}{\bibfnamefont{K.}~\bibnamefont{Lee}}, \bibnamefont{and}
  \bibinfo{author}{\bibfnamefont{A.~J.} \bibnamefont{Heeger}},
  \bibinfo{journal}{Nat. Photon.} \textbf{\bibinfo{volume}{3}},
  \bibinfo{pages}{297} (\bibinfo{year}{2009}).

\bibitem[{\citenamefont{Shuttle et~al.}(2008)\citenamefont{Shuttle, O'Regan,
  Ballantyne, Nelson, Bradley, de~Mello, and Durrant}}]{shuttle2008}
\bibinfo{author}{\bibfnamefont{C.~G.} \bibnamefont{Shuttle}},
  \bibinfo{author}{\bibfnamefont{B.}~\bibnamefont{O'Regan}},
  \bibinfo{author}{\bibfnamefont{A.~M.} \bibnamefont{Ballantyne}},
  \bibinfo{author}{\bibfnamefont{J.}~\bibnamefont{Nelson}},
  \bibinfo{author}{\bibfnamefont{D.~D.~C.} \bibnamefont{Bradley}},
  \bibinfo{author}{\bibfnamefont{J.}~\bibnamefont{de~Mello}}, \bibnamefont{and}
  \bibinfo{author}{\bibfnamefont{J.~R.} \bibnamefont{Durrant}},
  \bibinfo{journal}{Appl. Phys. Lett.} \textbf{\bibinfo{volume}{92}},
  \bibinfo{eid}{093311} (\bibinfo{year}{2008}),
  

\bibitem[{\citenamefont{Deibel et~al.}(2008)\citenamefont{Deibel, Baumann, and
  Dyakonov}}]{deibel2008b}
\bibinfo{author}{\bibfnamefont{C.}~\bibnamefont{Deibel}},
  \bibinfo{author}{\bibfnamefont{A.}~\bibnamefont{Baumann}}, \bibnamefont{and}
  \bibinfo{author}{\bibfnamefont{V.}~\bibnamefont{Dyakonov}},
  \bibinfo{journal}{Appl. Phys. Lett.} \textbf{\bibinfo{volume}{93}},
  \bibinfo{pages}{163303} (\bibinfo{year}{2008}).

\bibitem[{\citenamefont{Ju{\v{s}}ka et~al.}(2000)\citenamefont{Ju{\v{s}}ka,
  Arlauskas, Vili{\={u}}nas, and Ko{\v{c}}ka}}]{juska2000}
\bibinfo{author}{\bibfnamefont{G.}~\bibnamefont{Ju{\v{s}}ka}},
  \bibinfo{author}{\bibfnamefont{K.}~\bibnamefont{Arlauskas}},
  \bibinfo{author}{\bibfnamefont{M.}~\bibnamefont{Vili{\={u}}nas}},
  \bibnamefont{and}
  \bibinfo{author}{\bibfnamefont{J.}~\bibnamefont{Ko{\v{c}}ka}},
  \bibinfo{journal}{Phys. Rev. Lett.} \textbf{\bibinfo{volume}{84}},
  \bibinfo{pages}{4946} (\bibinfo{year}{2000}).

\bibitem[{\citenamefont{Langevin}(1903)}]{langevin1903}
\bibinfo{author}{\bibfnamefont{P.}~\bibnamefont{Langevin}},
  \bibinfo{journal}{Ann. Chim. Phys.} \textbf{\bibinfo{volume}{28}},
  \bibinfo{pages}{433} (\bibinfo{year}{1903}).

\bibitem[{\citenamefont{Pope and Swenberg}(1999)}]{pope1999}
\bibinfo{author}{\bibfnamefont{M.}~\bibnamefont{Pope}} \bibnamefont{and}
  \bibinfo{author}{\bibfnamefont{C.~E.} \bibnamefont{Swenberg}},
  \emph{\bibinfo{title}{Electronic Processes in Organic Crystals and Polymers}}
  (\bibinfo{publisher}{Oxford University Press}, \bibinfo{address}{USA},
  \bibinfo{year}{1999}), \bibinfo{edition}{2nd} ed.

\bibitem[{\citenamefont{Juska et~al.}(2006)\citenamefont{Juska, Arlauskas,
  Stuchlik, and Osterbacka}}]{juska2006}
\bibinfo{author}{\bibfnamefont{G.}~\bibnamefont{Juska}},
  \bibinfo{author}{\bibfnamefont{K.}~\bibnamefont{Arlauskas}},
  \bibinfo{author}{\bibfnamefont{J.}~\bibnamefont{Stuchlik}}, \bibnamefont{and}
  \bibinfo{author}{\bibfnamefont{R.}~\bibnamefont{Osterbacka}},
  \bibinfo{journal}{J. Non-Cryst. Sol.} \textbf{\bibinfo{volume}{352}},
  \bibinfo{pages}{1167} (\bibinfo{year}{2006}).

\bibitem[{\citenamefont{Deibel et~al.}(2009)\citenamefont{Deibel, Wagenpfahl,
  and Dyakonov}}]{deibel2009}
\bibinfo{author}{\bibfnamefont{C.}~\bibnamefont{Deibel}},
  \bibinfo{author}{\bibfnamefont{A.}~\bibnamefont{Wagenpfahl}},
  \bibnamefont{and} \bibinfo{author}{\bibfnamefont{V.}~\bibnamefont{Dyakonov}}
  (\bibinfo{year}{2009}), \bibinfo{note}{unpublished}.

\bibitem[{\citenamefont{B{\"a}ssler}(1993)}]{bassler1993}
\bibinfo{author}{\bibfnamefont{H.}~\bibnamefont{B{\"a}ssler}},
  \bibinfo{journal}{Phys. Stat. Sol. B} \textbf{\bibinfo{volume}{175}},
  \bibinfo{pages}{15} (\bibinfo{year}{1993}).

\bibitem[{\citenamefont{Rudenko and Arkhipov}(1982)}]{rudenko1982}
\bibinfo{author}{\bibfnamefont{A.}~\bibnamefont{Rudenko}} \bibnamefont{and}
  \bibinfo{author}{\bibfnamefont{V.~I.} \bibnamefont{Arkhipov}},
  \bibinfo{journal}{Phil. Mag. B} \textbf{\bibinfo{volume}{45}},
  \bibinfo{pages}{177} (\bibinfo{year}{1982}).

\bibitem[{\citenamefont{Arkhipov and Rudenko}(1982)}]{arkhipov1982}
\bibinfo{author}{\bibfnamefont{V.~I.} \bibnamefont{Arkhipov}} \bibnamefont{and}
  \bibinfo{author}{\bibfnamefont{A.}~\bibnamefont{Rudenko}},
  \bibinfo{journal}{Phil. Mag. B} \textbf{\bibinfo{volume}{45}},
  \bibinfo{pages}{189} (\bibinfo{year}{1982}).

\bibitem[{\citenamefont{Zaban et~al.}(2003)\citenamefont{Zaban, Greenshtein,
  and Bisquert}}]{zaban2003}
\bibinfo{author}{\bibfnamefont{A.}~\bibnamefont{Zaban}},
  \bibinfo{author}{\bibfnamefont{M.}~\bibnamefont{Greenshtein}},
  \bibnamefont{and} \bibinfo{author}{\bibfnamefont{J.}~\bibnamefont{Bisquert}},
  \bibinfo{journal}{Chem. Phys. Chem.} \textbf{\bibinfo{volume}{4}},
  \bibinfo{pages}{859} (\bibinfo{year}{2003}).

\end{thebibliography}
%\bibliographystyle{unsrt}

\end{document}